\documentstyle[12pt,epsfig]{article}
\begin{document}

\begin{center}
\large
Limits to Quantum Gravity Effects from \\ 
Observations of TeV Flares in Active Galaxies
\end{center}
\normalsize
\vskip .2in

\newcommand{\oxford}{$^{1}$ }
\newcommand{\dublin}{$^{2}$ }
\newcommand{\wash}{$^{3}$ }
\newcommand{\iowa}{$^{4}$ }
\newcommand{\maynooth}{$^{5}$ }
\newcommand{\purdue}{$^{6}$ }
\newcommand{\leeds}{$^{7}$ }
\newcommand{\caltech}{$^{8}$ }
\newcommand{\utah}{$^{9}$ }
\newcommand{\rtc}{$^{10}$ }
\newcommand{\whip}{$^{11}$ }
\begin{center}
S.D.~Biller,\oxford A.C.~Breslin,\dublin J.~Buckley,\wash 
M.~Catanese,\iowa M.~Carson,\dublin D.A.~Carter-Lewis,\iowa 
M.F.~Cawley,\maynooth
D.J.~Fegan,\dublin J.~Finley,\purdue J.A.~Gaidos,\purdue
A.M.~Hillas,\leeds  F.~Krennrich,\iowa R.C.~Lamb,\caltech
R.~Lessard,\purdue C.~Masterson,\dublin J.E.~McEnery,\utah
B.~McKernan,\dublin P.~Moriarty,\rtc
J.~Quinn,\whip H.J.~Rose,\leeds F.~Samuelson,\iowa 
G.~Sembroski,\purdue P.~Skelton,\leeds T.C.~Weekes\whip 
\\[2ex]
{\em \oxford Oxford University, Oxford, United Kingdom\\
\dublin University College, Dublin, Ireland\\
\wash Washington University, St. Louis, Missouri 63130\\
\iowa Iowa State University, Ames, Iowa 50011 \\
\maynooth Maynooth University, Maynooth, Ireland \\
\purdue Purdue University, West Lafayette, Indiana 47907\\
\leeds University of Leeds, Leeds, United Kingdom\\
\caltech California Institute of Technology, California 91125 \\
\utah University of Utah, Salt Lake City, UT 84112\\
\rtc Galway-Mayo Instutute of Technology, Galway, Ireland \\
\whip Whipple Observatory, Amado, Arizona 85645\\}

\end{center}

\vskip .2in

\begin{abstract}
\noindent We have used data from the TeV $\gamma$-ray flare associated with the active 
galaxy Markarian 421 observed on 15 May 1996 to place bounds
on the possible energy-dependence of the speed of light in the context
of an effective quantum gravitational energy scale. The possibility of 
an observable time dispersion in high energy radiation has recently 
received attention in the literature, with some suggestions that the 
relevant energy scale could be less than the Planck mass and perhaps as 
low as $10^{16}$GeV. The limits derived here indicate this energy scale 
to be in excess of $4\times10^{16}$GeV at the 95\% confidence level. 
To the best of our knowledge, this constitutes the first convincing limit
on such phenomena in this energy regime.
\end{abstract}

\vskip 0.3in
\begin{center}
{\large \em Submitted to Physical Review Letters}
\end{center}

\newpage

	It has recently been pointed out that many quantum gravity scenarios 
may result in an observable time dispersion for high energy radiation 
originating at large distances from the Earth \cite{Amelino} \cite{Garay} 
\cite{Gambini}. This would result from an effective energy-dependence to the 
velocity of light in vaccum owing to propagation through a gravitational 
medium containing quantum fluctuations on distance scales near the Planck 
length, $L_P \simeq 10^{-33}$cm, with timescales on the order of $1/E_P$, 
where $E_P$ is the Planck Mass ($\simeq10^{19}$GeV). In particular, it has 
been indicated \cite{Amelino} that different approaches to quantum gravity 
lead to a similar description of the first-order effects of such a time 
dispersion:
\begin{equation}
\Delta t \simeq \xi \frac{E}{E_{QG}} \frac{L}{c}
\end{equation}
where $\Delta t$ is the time delay relative to the standard energy-independent 
speed of light, $c$; $\xi$ is a model-dependent factor of order 1;
$E$ is the energy of the observed radiation; $E_{QG}$
is the assumed energy scale for quantum gravitational effects
which can couple to electromagnetic radiation;
and $L$ is the distance over which the radiation has propagated.
While $E_{QG}$ is generally assumed to be on the order of $E_P$,
recent work within the context of string theory suggests that the onset
of noticeable quantum gravitational effects may correspond to a characteristic 
energy scale smaller than the Planck mass and perhaps as low as 
$10^{16}$GeV \cite{Witten}. Thus, any experimental probe of such scales 
or higher would be of great interest.

	In a recent paper \cite{Amelino} it was suggested that 
$\gamma$-ray bursts (GRBs) could provide a natural way to test such predictions 
owing to the short duration, high energies and the apparent cosmological 
origin of at least some of these bursts. Based on current data, these authors 
indicate that if {\bf 1)} time structure on the scale of 0.01 seconds or smaller 
can be established for energies $\sim200$keV and {\bf 2)} an association of
such a burst can be made with an object possessing a redshift of order 1, 
energy scales of $E_{QG} \sim 10^{16}$GeV could be probed. Unfortunately, 
establishing the distance of any particular GRB from earth has proven to be 
non-trivial, with only a handful positively associated with optical 
counterparts. Also, some of the highest energies seen from GRBs are associated with 
an ``afterglow'' which seems to occur over much longer timescales than
the initial burst. However, more stringent and robust limits to $E_{QG}$ 
can already be set based instead on the rapidly rising TeV flares seen to
occur in active galaxies.

	The Whipple Observatory $\gamma$-ray telescope, located in Arizona, 
detects the \v{C}erenkov light generated by electromagnetic cascades resulting 
from the interaction of high-energy $\gamma$-rays in the atmosphere. Images 
taken of such cascades
are used to discriminate backgrounds and derive energies of the primary
$\gamma$-rays in the regime above $\sim$250 GeV. To date, three extragalactic 
sources, all active galaxies of the blazar class, have been identified as
emitters of TeV radiation \cite{Punch} \cite{Quinn} \cite{Catanese1}. 
Two of these, Markarian 421 and Markarian 501,
produce particularly strong emission with energy spectra approximated by an
$\sim E^{-2.5}$ power law (although Markarian 501 shows evidence for additional
curvature) between energies of 300 GeV and 10 TeV \cite{Krennrich}. These same 
sources have also exhibited dramatic changes in flux level on timescales 
ranging from minutes to days. On several occasions, such variations have been 
simultaneously studied and correlated with x-ray, UV and optical measurements 
\cite{Buckley} \cite{Catanese2}. 

	The most rapid flare observed thus far was seen from Markarian 421 
on 15 May 1996 \cite{Gaidos}. This data is shown in figure 1, where the 
excess rate of $\gamma$-ray selected events above a threshold of 350 GeV 
is binned in intervals of 280 seconds duration, as it appeared in the 
original publication of this observation. To avoid 
confusion (and potential bias), we will retain this same binning for the 
current analysis. The doubling time of the flare is less than 15 minutes, 
although variability is apparent on the scale of the binning at the 99\% 
confidence level. Because of the rapidly falling energy spectrum, the
$\gamma$-ray data is dominated by events near the triggering threshold.
Thus, the peak of the flare is almost entirely defined by events with
$\gamma$-ray energies less than 1 TeV, as shown at the top of figure 2.
In the 280 second interval corresponding to this peak,
4 events with $\gamma$-ray energies in excess of 2 TeV have been 
identified, whereas no such events could be identified in either of
the 2 adjacent intervals (figure 2, bottom).
The probability for this to occur by chance is less than 5\%. Hence, at 
the 95\% confidence level, emission above 2 TeV appears to keep in step 
with emission below 1 TeV for variability timescales less than 280 seconds. 
The redshift of Markarian 421 is 0.031, which translates 
to $1.1\times10^{16}$ light-seconds for an assumed Hubble constant of 
85 km/s/Mpc. From equation 1, this leads to lower bound on $E_{QG}$/$\xi$ 
of $4\times10^{16}$GeV.

	We note that an earlier limit on the energy-dependence of the speed
of light, which would be more restrictive than that given here, 
had been derived from the possible ultra-high-energy detection of anomalous
pulsed emission from Her X-1 in 1986 \cite{Haines}. However, more recent 
analyses and the lack of further such detections suggests that the 
interpretation of that observation as a statistical fluctuation is
not an unreasonable one \cite{Biller}.
We therefore believe that the limit presented in this paper represents
the most credible and stringent bound thus far obtained. 

	The next generation of proposed ground-based instruments, such 
as VERITAS and HESS, will feature multi-telescope systems with much 
improved sensitivity, energy coverage and resolution, along with the 
ability to track candidate sources of flares more continuously using 
dedicated telescopes. This will allow for both a more detailed study
of the time structure of currently known TeV sources and the
prospect of discovering and studying more distant objects. It is
therefore reasonable to expect to probe $E_{QG}$ to even higher energies
in the near future from further studies of TeV flares. 
As has already been
pointed out \cite{Amelino}, the distinctive dependence of the shortest 
observable variability timescale on both energy and source distance for quantum 
gravitational dispersion should allow source-specific effects to be 
distinguished. Thus, future TeV studies could conceivably provide convincing
evidence for quantum gravity, particularly if the resulting time-dispersion 
effects are associated with characteristic energy scales less than the Planck 
mass. We hope that this prospect, in addition to the bounds derived here, will
encourage more detailed predictions of such phenomena to be calculated in
the context of specific quantum gravity frameworks.

	We would like to thank Subir Sarkar at Oxford for several useful 
discussions on this topic. This work has been supported in part by PPARC, 
Forbairt, the US Department of Energy, NASA and the Smithsonian Institution.

\begin{figure}[p]
\setlength{\epsfxsize}{0.85\textwidth}
\leavevmode
\epsfbox{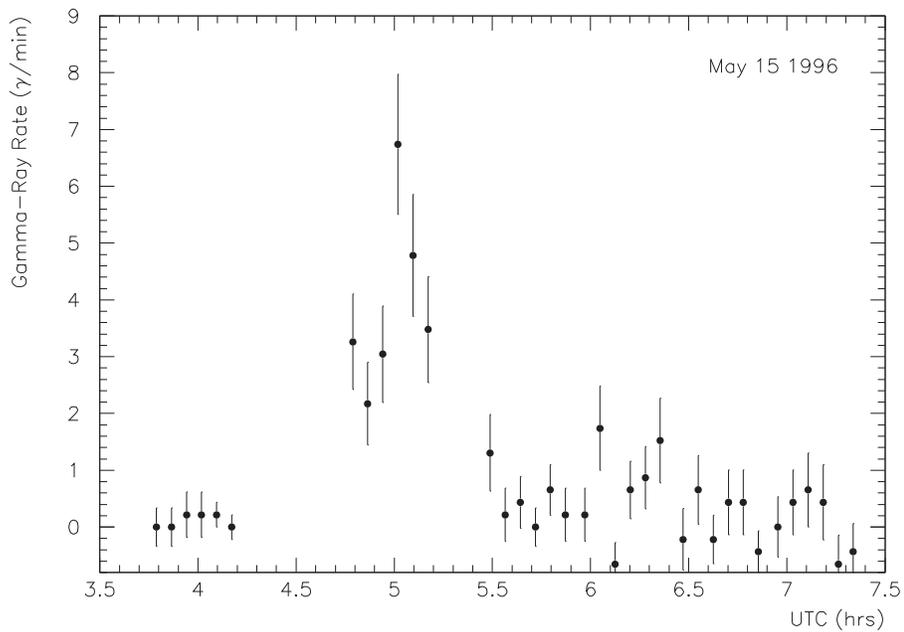}
\caption[Flare from Markarian 421]
{TeV $\gamma$-ray flare from Markarian 421 observed on
15 May 1996 by the Whipple $\gamma$-ray observatory.
The rate of excess $\gamma$-ray selected events is binned 
in intervals of 280 seconds.
(taken from reference \cite{Gaidos}) }
\label{may_flare} \end{figure}

\begin{figure}[p]
\setlength{\epsfxsize}{0.85\textwidth}
\leavevmode
\epsfbox{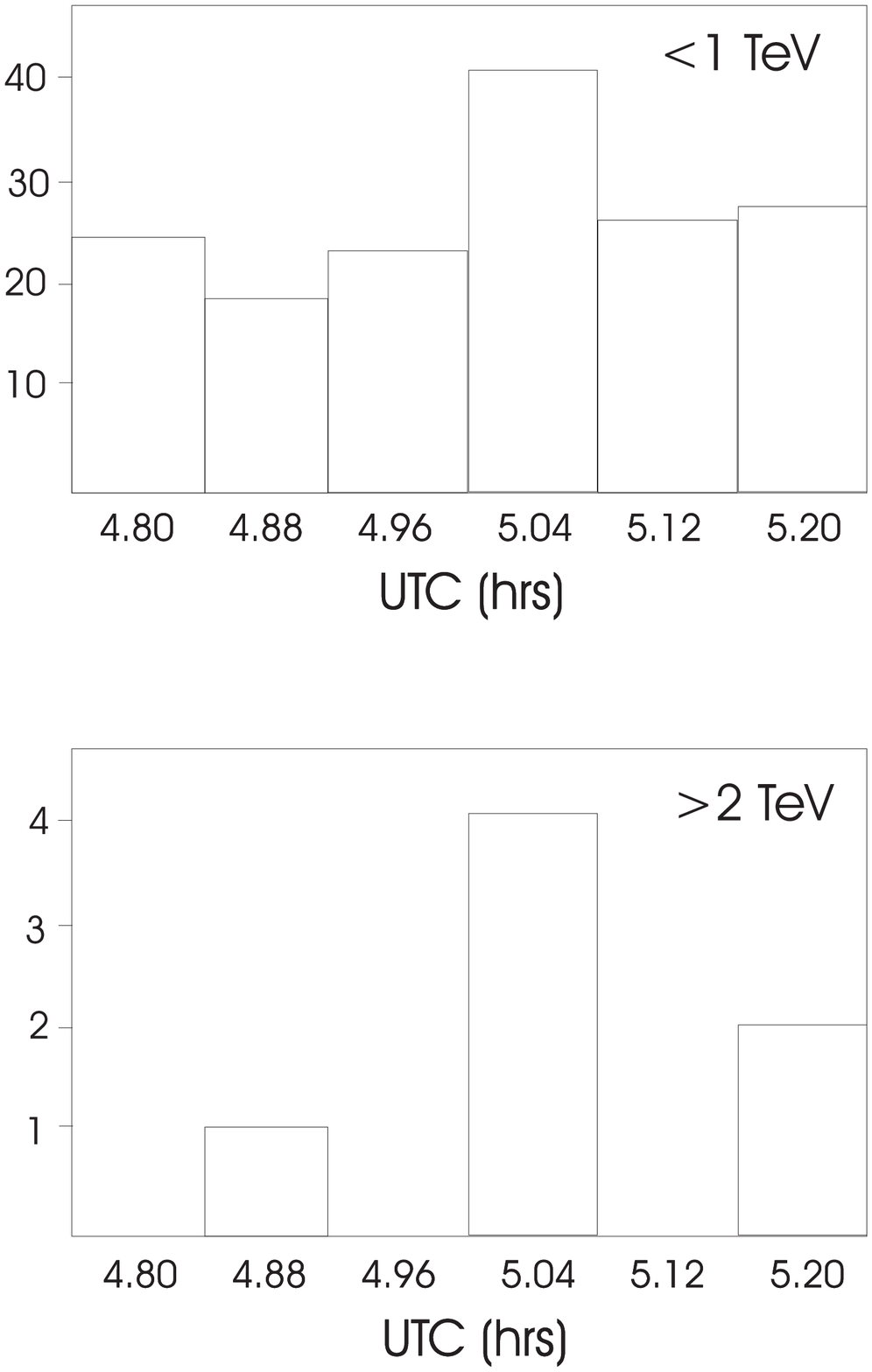}
\caption[Flare from Markarian 421]
{Total number of $\gamma$-ray selected events occurring in
each 280 second interval near the peak of the 15 May 1996 flare
from Markarian 421. The top plot consists of events with
$\gamma$-ray energies less than 1 TeV, whereas the bottom plot
is for energies greater than 2 TeV.}
\label{flare_comp} \end{figure}

\end{document}